%% file: AMSP_TMSP.tex
\documentclass[graybox]{svmult}

\usepackage{mathptmx}       
\usepackage{helvet}         
\usepackage{courier}        
\usepackage{type1cm}        
\usepackage{mathrsfs}
\usepackage{amsmath}
\usepackage{amssymb}
\usepackage{makeidx}         
\usepackage{graphicx}        
\usepackage{multicol}        
\usepackage[bottom]{footmisc}

\def\ltsima{$\; \buildrel < \over \sim \;$}
\def\lsim{\lower.5ex\hbox{\ltsima}}
\def\gtsima{$\; \buildrel > \over \sim \;$}
\def\gsim{\lower.5ex\hbox{\gtsima}}
\def\mdot {\dot M}
\newcommand{\be}{\begin{equation}}
\newcommand{\en}{\end{equation}}

\def\msole {~M_{\odot}}

\makeindex             

\begin{document}

\title*{Accreting pulsars: mixing-up accretion phases in transitional systems}
\author{Sergio Campana and Tiziana Di Salvo}
\institute{INAF - Osservatorio Astronomico di Brera, Via E. Bianchi 46, I-23807, Merate (LC), Italia, \email{sergio.campana@brera.inaf.it}\\
Dipartimento di Fisica e Chimica, Universit\`a di Palermo, via Archirafi 36, I-90123, Palermo, Italia, \email{tiziana.disalvo@unipa.it}}

\maketitle

\abstract{In the last 20 years our understanding of the millisecond pulsar population
changed dramatically. Thanks to the  large effective
area and good time resolution  of the NASA X--ray observatory
Rossi X-ray Timing Explorer, we discovered that neutron stars
in Low Mass X-ray Binaries (LMXBs) spins at frequencies between 200
and 750 Hz, and indirectly confirmed the recycling scenario, according
to which neutron stars are spun up to millisecond periods during the
LMXB-phase. 
In the meantime, the continuous discovery of rotation-powered millisecond
pulsars in binary systems in the radio and gamma-ray band (mainly with
the Fermi Large Area Telescope) allowed us to classify these
sources into two ``spiders" populations, depending on
the mass of their companion stars: Black Widow pulsars,
with very low-mass companion stars, and Redbacks, with
larger mass companion stars possibly filling their Roche lobes without accretion
of matter onto the neutron star.
It was soon regained that millisecond pulsars in short orbital
period LMXBs are the progenitors of the spider populations of
rotation-powered millisecond pulsars, although a direct link between
accretion-powered and rotation-powered millisecond pulsars was still
missing.
In 2013 the ESA X-ray observatory XMM-Newton spotted the X--ray outburst
of a new accreting millisecond pulsar (IGR J18245-2452)
in a source that was previously classified as a radio millisecond pulsar,
probably of the Redback type. Follow up observations of the source when
it went back to X--ray quiescence showed that it was able to swing between
accretion-powered to rotation-powered pulsations in a relatively short
timescale (few days), promoting this source as the direct link between
the LMXB and the radio millisecond pulsar phases. Following discoveries
showed that there exists a bunch of sources which alternates X--ray activity
phases, showing X--ray coherent pulsations, to radio-loud phases, showing
radio pulsations, establishing a new class of millisecond pulsars, the
so-called transitional millisecond pulsars.
In this review we describe these exciting discoveries and the properties
of accreting and transitional millisecond pulsars, highlighting what we
know and what we have still to learn about in order to fully understand the
(sometime puzzling) behaviour of these systems and their evolutive connection. }

\section{The links in the chain: how a neutron star becomes a millisecond pulsar}
\label{sec:1}

\subsection{The recycling scenario: the evolutionary path leading to the formation 
of millisecond pulsars}
\label{msp}

A millisecond pulsar (hereafter MSP) is a fast rotating, weakly magnetised 
neutron star. A weak magnetic field for a neutron star means a magnetic field
of $\sim 10^7-10^8$ G, that is several orders of magnitude higher than the 
strongest magnetic fields that can be produced on Earth laboratories (the
highest magnetic field strength created on Earth is $\sim 9 \times 10^5$ 
G\footnote{https://phys.org/news/2011-06-world-strongest-magnetic-fields.html}), 
but weak with respect to the magnetic field of a newly born neutron 
star (which is $\gtrsim 10^{11}-10^{12}$ G). 
MSPs have spin periods in the range 1--10 ms, corresponding to spin 
frequencies above 100 Hz. MSPs were first discovered in the radio band, with 
the detection of periodic radio pulses. The first discovered MSP is PSR B1937+21 
\cite{Backer1982}, spinning roughly 641 times a second; this is to date the 
second fastest-spinning MSP among the $\sim 300$ that have been discovered 
so far. PSR J1748--2446ad, discovered in 2005 \cite{Hessels2006}, is the 
fastest-spinning pulsar currently known, spinning at 716 Hz. 
These millisecond spinning neutron stars are extreme physical objects: 
general and special relativity are fully in action, since their surfaces, 
attaining speeds close to one fifth of the speed of light, are located 
extremely close to their Schwarzschild radius. In addition electro-dynamical forces, 
caused by the presence of huge surface magnetic fields of several hundred million 
Gauss, display their spectacular properties accelerating electrons up to such 
energies to promote pair creation in a cascade process responsible for the 
emission in the radio and $\gamma$-ray bands. The rotational energy is swiftly 
converted and released into electromagnetic power which, in some cases, causes 
the neutron star to outshine with a luminosity of hundreds Suns. 

Standard radio pulsars are usually isolated objects, with relatively high 
magnetic field strengths ($\gtrsim 10^{11}$ G) and relatively long spin periods 
($\gtrsim 0.1$ s). 
Neutron star magnetic fields are probably the relic magnetic field of the 
progenitor star that is enhanced by ``flux freezing", or conservation of the 
original magnetic flux, when the core of the progenitor star collapses to form 
a neutron star. 
Moreover, as the core of a 
massive star ($\ge 8\, M_\odot$) is compressed and collapses into a neutron 
star, it retains most of its angular momentum. But, because it has only a 
tiny fraction of the radius of its progenitor star, a neutron star is formed 
with very high rotation speed. 
An interesting example of a recently formed pulsar is 
the Crab pulsar, the central star in the Crab 
Nebula, a remnant of the supernova SN 1054, which exploded in the year 1054,
less than a thousand years ago, that shows a spin period of 33 ms and a magnetic
field strength of $B \gtrsim 4 \times 10^{12}$ G.

The strong magnetic field of the newly born neutron star and the high 
rotational velocity at its surface generate a strong Lorentz force
resulting in the acceleration of protons and electrons on the star 
surface and the creation of an electromagnetic beam emanating from the 
poles of the magnetic field, which is responsible for the observed pulsed 
emission. In rotation-powered pulsars, the energy of the beam comes from
the rotational energy of the pulsar, which therefore starts to spin-down.
At zero-order the pulsar behaves as a rotating magnetic dipole which 
emits energy according to the Larmor formula (see Jackson's Classical 
Electrodynamics):
\begin{equation}
\label{Larmor}
P_{\rm rad} = {2 \over 3} { (\ddot{m}_\bot )^2 \over c^3} = 
{2 \over 3} { m_\bot^2 \Omega^4 \over c^3} = {2 \over 3
c^3} ( B R^3 \sin \alpha)^2 \biggl( {2 \pi \over P} \biggr)^4~,
\end{equation}
where $m_\bot = B R^3 \sin \alpha$ is the component of the magnetic dipole 
moment perpendicular to the rotation axis, $B$ and $R$ are the surface
magnetic field and the neutron star radius, respectively, $\alpha$ is the 
angle between the rotation axis and the magnetic dipole axis, $\Omega$ is the 
spin angular frequency of the neutron star and $P$ its spin period. 
The pulsar, therefore, gradually slows down at a rate that is higher for
stronger magnetic fields and faster spins. If the magnetic field strength 
does not change significantly with time, we can estimate a pulsar's age  
from its spin period and the spin-down rate, by assuming that the pulsar's 
initial period $P_0$ was much shorter than the current period: 
\begin{equation}
\label{Age}
\tau \equiv { P \over 2 \dot{P}}.
\end{equation}
This is the timescale necessary to bring the pulsar from its initial spin
period $P_0$ to its actual period at the observed spin-down rate, and is 
called characteristic age of the pulsar. Indeed the characteristic age of
the Crab is $\sim 2.5$ kyr.

Soon after their discovery it became clear that MSPs are old neutron stars, 
with relatively weak magnetic fields, of characteristic ages comparable to
the age of the Universe; these were therefore recognised as a different class
of objects, called {\it recycled pulsars}. 
Binary systems that can host an old neutron star and may be responsible 
for the recycling are the so-called Low Mass X--ray Binaries (hereafter LMXBs), 
in which an old, weakly magnetised, neutron star accretes matter from a 
low-mass (less that $1\, M_\odot$) companion star. In these systems, 
matter transferred from the companion star enters the Roche lobe of the 
neutron star through the inner Lagrangian point and releases a large amount 
of gravitational energy before falling onto the neutron star. 
These systems may be up to five orders of 
magnitude more luminous than the Sun, and the temperature matter reaches 
close to the neutron star ranges from few keV up to a hundred keV, making 
these systems the brightest Galactic sources in the X--ray band. Because 
of the small size of the companion star, these systems are also quite compact,
with orbital periods ranging from several minutes up to one day. For this
reason, matter leaving the companion star, has a high specific angular 
momentum and cannot fall directly onto the neutron star. Besides an 
accretion disc is formed, where matter rotates with Keplerian velocities
and looses energy until it reaches the innermost part of the system, close
to the neutron star. When this matter accretes onto the neutron star 
surface it has relativistic velocities (up to half the velocity of light),
and is able to efficiently accelerate the neutron star up to millisecond
periods. Depending on the Equation of State (EoS) of ultra-dense matter,
$0.1-0.2\, M_\odot$ are sufficient to spin up a weakly magnetised neutron 
star to millisecond periods \cite{Burderi1999}. During this phase, 
because of the accretion of matter and angular momentum, the neutron star 
accumulates an extraordinary amount of mechanical rotational energy, 
up to 1\% of its whole rest-mass energy. 

\subsection{Problems and confirmation of the recycling scenario}
\label{recycl}

The recycling scenario described above (see e.g.\ \cite{Bhattacharya1991})
establishes therefore a clear evolutive link between
LMXBs and radio MSPs, with the former being the progenitors of the latter. 
However, till the end of nineties, there was no observational evidence confirming
this evolutive scenario, since there was no evidence that LMXBs could host
fast rotating neutron stars. In fact, despite thoroughly searched, no LMXB
was found to show coherent pulsations, therefore unveiling its spin frequency.
The fact that the large majority of LMXBs does not show coherent pulsations
is still a problem. Several explanations have been invoked to interpret this
fact, but none of them is fully satisfactory (see also \cite{Patruno2012a}, 
and references therein, for further discussion of this issue). 
One possibility is that the magnetic axis is aligned with the rotation axis 
(e.g.\ \cite {Ruderman1991}), but this is excluded by the fact that MSPs, the 
descendants, show pulsations and therefore the magnetic and rotational axes 
are not aligned in these systems. 
Another possibility is that the magnetic field is not strong enough to 
channel the accreting matter to the neutron star magnetic poles or 
that optically thick matter around the neutron star may smear the coherent 
pulsations (e.g.\ \cite{Brainerd1987}). However, the large majority of LMXBs are 
transient systems, 
showing large variation in luminosity, and going through soft (optically 
thick) X--ray spectra at high luminosity and hard (optically thin) X--ray 
spectra at low luminosity. 
During these stages the accretion rate decreases, the accreting matter 
becoming optically thin, allowing in principle the detection of X--ray 
pulsations at millisecond periods (e.g.\ \cite{Gogus2007}). 
Other two possibilities are that the neutron star magnetic field is
buried by long phases of accretion (e.g.\ \cite{Romani1990,Cumming2001}) 
or that the coherent pulsations are very weak, beyond the sensitivity of 
current X--ray observatories.

Another problem of the recycling scenario was the lack of the observational 
link between LMXBs and MSPs, i.e.\ the lack of a system behaving like a LMXB 
during X--ray active phases and like a radio MSP during X--ray quiescence, 
when presumably the accretion rate goes down and the radio pulsar mechanism 
can switch on. 
However, lack of evidence does not mean evidence of lack, and both these
problems were recently solved. In particular, in 1998 coherent pulsations
were discovered for the first time in a transient LMXB (SAX J1808.4-3658,
\cite{Wijnands1998}), and in 2013 the long-sought-for missing 
link between LMXBs and MSPs was finally found (IGR J18245--2452, a.k.a. M28I,
\cite{Papitto2013a}). In the next section we describe these discoveries and
the related ones, and give the basic observational characteristics of
these new classes of systems.

\begin{table}
\caption{Accreting X--ray Pulsars in Low Mass X--ray Binaries.}
\scriptsize
\begin{center}
\begin{tabular}{lllllll}
\hline
\hline
Source & $\nu_{s}/P$ & $P_{\rm orb}$ & $f_{x}$  & $M_{c,min}$  & Companion  & Ref.\\
 & (Hz)/(ms) & (hr) & ($M_{\odot}$) & ($M_{\odot}$) &   Type & \\
\hline
\textbf{Accreting Millisecond Pulsars}\\
\hline
XSS  J12270--4859 & 593 (1.7) & 6.91 & $3.9\times 10^{-3}$ & 0.27 & MS & \cite{Roy2015,deMartino2014}\\
PSR J1023+0038 & 592 (1.7) & 4.75 & $1.1\times 10^{-3}$ & 0.20 & MS & \cite{Archibald2009,CotiZelati2014}\\
Aql X-1     & 550 (1.8) &  18.95 & $1.4\times 10^{-2}$ & 0.56 & MS & \cite{Casella2008,MataSanchez2017}\\
Swift J1749.4--2807 & 518 (1.9) & 8.82 & $5.5\times 10^{-2}$ & 0.59 & MS & \cite{Altamirano2011,DAvanzo2011}\\
SAX J1748.9--2021  & 442 (2.3) & 8.77 &  $4.8\times 10^{-4}$ & 0.1  & MS& \cite{Altamirano2008,Cadelano2017}\\
IGR J17498--2921 & 401 (2.5) & 3.84 & $2.0\times10^{-3}$ & 0.17 & MS & \cite{Papitto2011b}\\
XTE J1814--338  & 314  (3.2) & 4.27 & $2.0\times 10^{-3}$ & 0.17 & MS  & \cite{Markwardt2003,Wang2017}\\
IGR J18245--2452 & 254 (3.9) & 11.03 & $2.3\times10^{-3}$ & 0.17 & MS & \cite{Papitto2013a}\\
IGR J17511--3057 & 245 (4.1) &  3.47 & $1.1\times 10^{-3}$ & 0.13  & MS & \cite{Papitto2010}\\
\hline
IGR J00291+5934    & 599  (1.7) & 2.46 & $2.8\times 10^{-5}$ & 0.039  &  BD & \cite{Galloway2005}\\
SAX J1808.4--3658 & 401 (2.5) &  2.01 & $3.8\times 10^{-5}$ & 0.043  & BD &  \cite{Wijnands1998,Wang2013}\\
HETE J1900.1--2455& 377  (2.7) &   1.39 & $2.0\times 10^{-6}$ & 0.016  & BD  & \cite{Kaaret2006,Elebert2008}\\
\hline
XTE J1751--305 & 435 (2.3) &  0.71 & $1.3\times 10^{-6}$ & 0.014  & He WD &   \cite{Markwardt2002,DAvanzo2009}\\
MAXI J0911--655 & 340 (2.9) & 0.74 & $6.2\times 10^{-6}$ & 0.024  & He WD? &   \cite{Sanna2017a}\\
NGC6440 X--2 & 206 (4.8) & 0.95 & $1.6\times 10^{-7}$ & 0.0067 & He WD & \cite{Altamirano2010}\\
Swift J1756.9--2508  & 182  (5.5) &  0.91 &  $1.6\times 10^{-7}$ & 0.007  & He WD &  \cite{Krimm2007}\\
IGR J16597--3704 & 105 (9.5) & 0.77 & 1.2$\times 10^{-7}$ & 0.006 & He WD & \cite{Sanna2018}\\
\hline
XTE J0929--314  & 185  (5.4) & 0.73 & $2.9\times 10^{-7}$ & 0.0083  & C/O WD  & \cite{Galloway2002,Giles2005}\\
XTE J1807--294  & 190  (5.3) & 0.67 & $1.5\times 10^{-7}$ & 0.0066  & C/O WD  & \cite{Campana2003,DAvanzo2009} \\
\hline
IGR J17062--6143 & 164 (6.1) & $>0.28$& -- & -- & -- &  \cite{Strohmayer2017}\\
\hline
\hline
\end{tabular}\\
\end{center}
$\nu_{s}$ is the spin frequency, $P_{b}$ the orbital period, $f_{x}$
is the X--ray mass function, $M_{c,min}$ is the minimum companion mass
for an assumed NS mass of 1.4 M$_\odot$. 
The companion types are: WD = White Dwarf, BD= Brown Dwarf, MS = Main Sequence, He Core = Helium Star.\newline
$^{b}$ Binary with parameters that are still compatible with an intermediate/high mass donor.\newline
Adapted and updated from \cite{Patruno2012a}.
\label{tab:lmxbs}
\end{table}

\section{Millisecond pulsars in LMXBs and their properties}
\label{sec:2}

\subsection{The discovery of a new class of fast spinning neutron star}
\label{discovery}

The situation dramatically changed in 1996, when the NASA observatory 
Rossi X--ray Timing Explorer (RXTE) was launched. RXTE was the first
X--ray observatory coupling a large effective area (the Proportional Counter
Array, PCA, had a total collecting area of $\sim 6,500$ cm$^2$) and good time
resolution (up to $1\, \mu$s), hence suitable for the search of fast time 
variability. In 1996 quasi-coherent pulsations were detected in RXTE 
observations of the LMXB 4U 1728--34 at a frequency of $\sim 363$ Hz, 
with amplitudes (rms) of $2.5\%-10\%$ during six of the eight type-I bursts
present in the observation \cite{Strohmayer1996}. The pulsations 
during these bursts showed frequency drifts of 1.5 Hz during the first 
few seconds but became effectively coherent during the burst decay. 
The 363 Hz pulsations were interpreted as rotationally induced modulations 
of inhomogeneous burst emission, and were considered the first compelling 
evidence for a millisecond spin period in a LMXB.  

The direct evidence for the presence of a fast-spinning neutron star in a LMXB 
arrived two years later, in 1998, when observations performed with RXTE led 
to the discovery of the first millisecond pulsar in a LMXB, SAX J1808.4-3658. 
This transient LMXB, first observed by the Wide Field Camera (WFC) on board
the X--ray satellite BeppoSAX, shows coherent pulsations with a period of 
2.5 ms and an orbital period of 2.01 hr (\cite{Wijnands1998}, 
\cite{Chakrabarty1998}). For almost four years, SAX J1808.4-3658 was 
considered as a rare object in which some peculiarity of the system allowed for
the detection of the neutron star spin. However, in the last 20 years, other 
19 accreting millisecond pulsars have been discovered, the two most recent 
ones discovered in 2017 (\cite{Sanna2017a,Strohmayer2017}). 
All of them show coherent pulsations with periods in the range 1.7--6.0 ms
(up to 60 ms if we also include the enigmatic LMXB IGR J17480--2446 recently
discovered in the Globular Cluster Terzan 5; \cite{Papitto2011a}), 
and all of them are found in compact systems, with orbital periods in the 
range 40 min to $\sim 10$ hr (with the exception is Aql X-1, one of the so-called 
intermittent millisecond pulsars, which has an orbital period of 19 hr). 
Hence, very low mass donors, $\le 0.2\, M_\odot$, are usually preferred. 
Another common feature of these systems is that all of them are transients, 
spending most of the time in a quiescent X--ray state; on occasions they show 
X--ray outbursts with moderate peak luminosities in the range $10^{36}-10^{37}$ 
erg s$^{-1}$. It was clear that we were facing a new class of astronomical objects, 
the so-called Accreting Millisecond X--ray Pulsars (hereafter AMSPs) that could 
constitute the bridge between the accretion-powered (LMXBs) and the 
rotation-powered (MSPs) neutron star sources.

\subsection{Peculiar behaviours and intermittent pulsations}
\label{intermittent}

Most (if not all) of the AMSPs are transient, as the vast majority of LMXBs 
in general. They spend most of time in quiescence with very low X--ray
luminosity ($\lesssim 10^{31}-10^{33}$ erg s$^{-1}$) and sometimes they show X--ray 
outbursts (reaching luminosities of $\sim 10^{36}-10^{37}$ erg s$^{-1}$) usually 
lasting from few days to $\le 3$ months. The shortest outburst recurrence time is one 
month for the globular cluster source NGC 6440 X-2, with an outburst duration 
of less than $4-5$ days, whereas the longest outburst, from HETE J1900.1--2455, 
has lasted for $\sim 10$ years (up to late 2015 when the source returned to 
quiescence, \cite{Degenaar2017}). However, most of these systems have shown 
just one X--ray outburst during the last 20 years. 
The most regular among recurrent AMSPs, and therefore the best studied of these 
sources, is the first discovered AMSP, SAX J1808.4-3658, which has shown an 
X--ray outburst every 1.6--3.5 years, the latest one occurred in 2015 (\cite{Patruno2017},
\cite{Sanna2017b}). The outburst light curve is characterised by a fast rise 
(on a couple of days timescale), a slow exponential decay (with a timescale 
of $\sim 10$ days) followed by a fast decay (with a timescale of $\sim 2$ days).
After the end of the main outburst, usually a flaring activity, called 
{\it reflares}, is observed, with a quasi-oscillatory behaviour and a 
variation in luminosity of up to three orders of magnitude on timescales 
$\sim 1-2$ days. Moreover a strong $\sim 1$ Hz oscillation is observed to 
modulate the reflares. A similar behaviour was also observed in the AMSP
NGC 6440 X-2 (see e.g.\ \cite{Patruno2013}). 
The reflaring behaviour has no clear explanation. \cite{Patruno2016} 
proposed a possible explanation in terms of either a strong propeller with 
a large amount of matter being expelled from the system or a trapped 
(dead) disc truncated at the co-rotation radius. 

Another peculiar behaviour is the intermittency of the pulsations, important 
because it could bridge the gap between non-pulsating LMXBs and AMSPs. In 
2005, the seventh discovered AMSP, HETE J1900.1--2455, went into X--ray
outburst and showed X--ray pulsations at 377 Hz, with an orbital period 
of 1.39 hr \cite{Kaaret2006}. Contrary to the usual behaviour of AMSPs,
this outburst lasted for about 10 years. After the first 20 days of the 
outburst, the pulsations became intermittent 
for about 2.5 years. After that the pulsed fraction weakened with stringent
upper limits ($\le 0.07\%$, \cite{Patruno2012b}). The most puzzling behaviour in
this sense, was observed in the LMXB Aql X-1, which showed coherent X--ray
pulsations, discovered in 1998 RXTE archival data, that appeared in only 
one $\sim 150$ s data segment out of a total exposure time of 1.5 Ms from more 
than 10 years of observations \cite{Casella2008}. The third intermittent
pulsar is SAX J1748.9--2021, where pulsations were detected sporadically in 
several data segments and in three out of four outbursts observed by
the source (\cite{Patruno2009a}, see also \cite{Sanna2016} reporting on
the 2015 outburst of the source). Interestingly, these AMSPs may have 
a long term average mass accretion rate higher with respect to the other AMSPs.
To explain this behaviour, it has been proposed that a screening of the 
neutron star magnetic field by the accreting matter weakens its strength 
by orders of magnitude on timescale of few hundred days, so that it is
less effective in truncating the accretion disc and channel matter to the
magnetic poles \cite{Patruno2012b}. However, it is not clear if this hypothesis 
can explain all the phenomenology and more observations and theoretical
efforts are needed to reach a satisfactory explanation. 

A detailed review of most of the phenomenology of AMSPs can be found in
\cite{Patruno2012a}. In the following we will give an overview of
the most debated issues on these systems, with particular attention 
to aspect regarding their evolution and their connection to 
rotation-powered MSPs.

\subsection{Accretion torques and short-term spin variations}
\label{torques}

Accretion torque theories can be tested studying the spin variations of
AMSPs during accretion states. These studies can provide valuable information 
on the mass accretion rate and magnetic field of the neutron star in 
these systems, as well as their spin evolution. 
An open question is whether these accreting pulsars are spinning up 
during an outburst and spinning down in quiescence as predicted by 
the recycling scenario. 
Coherent timing has been performed on several sources of the sample,
with controversial results. Although some AMSPs show pulse phase delays 
distributed along a second order polynomial, indicating an almost constant 
spin frequency derivative, other sources show strong timing noise which can 
hamper any clear measurement of the spin derivative. In fact, the phase delays 
behaviour as a function of time in these sources is sometimes quite complex 
and difficult to interpret, since phase shifts, most probably related to
variations of the X--ray flux, are sometimes present.  

The first AMSP for which a spin derivative has been measured is the 
fastest spinning ($\sim 599$ Hz, in a 2.46 hr orbit) among these sources, 
IGR J00291+5934. It is now generally accepted that this source shows
spin up at a rate of $\sim (5-8) \times 10^{-13}$ Hz s$^{-1}$ (\cite{Falanga2005}, 
\cite{Patruno2010}). \cite{Burderi2006} have attempted to fit
the phase delays vs.\ time with physical models taking into account the 
observed decrease of the X--ray flux as a function of time during the 
X--ray outburst, with the aim to get a reliable estimate of the mass accretion 
rate onto the compact object. In the hypothesis that the spin-up of the 
source is caused by the accretion of matter and angular momentum from 
a Keplerian accretion disc, the mass accretion rate, $\dot M$, onto the 
neutron star can be calculated by the simple relation: 
$2 \pi I \dot \nu = \dot M (G M_{NS} R)^{1/2}$, where $I$ is the 
moment of inertia of the neutron star, $\dot \nu$ the spin frequency 
derivative, $G$ the gravitational constant, $M_{NS}$ the neutron star 
mass, $R$ the accretion radius, and $(G M_{NS} R)^{1/2}$ the Keplerian
specific angular momentum at the accretion radius. 
Because the X--ray flux, which is assumed to be a good tracer of the mass 
accretion rate, is observed to decrease along the outburst, this has 
to be included in the relation above in order to obtain the correct value 
of the mass accretion rate at the beginning of the outburst as well as 
its temporal evolution. Note that the accretion radius also depends 
on the mass accretion rate, $R \propto \dot M^{-\alpha}$, where $\alpha$
is usually assumed to be $2/7$ (e.g.\ \cite{Ghosh1978}), and therefore 
varies with time. Fitting the phase delays in this way, the spin
frequency derivative at the beginning of the outburst results to be
$\dot \nu \sim 1.2(2) \times 10^{-12}$ Hz s$^{-1}$, and the lower limit to the 
mass accretion rate at the beginning of the outburst, corresponding to 
$\alpha = 0$, is $\dot M_{-10} = 5.9 \dot \nu_{-13} I_{45} m^{-2/3} =
70 \pm 10$, where $\dot M_{-10}$ is the mass accretion rate in units of 
$10^{-10}\, M_\odot$ yr$^{-1}$, $\dot \nu_{-13}$ is the spin frequency derivative 
in units of $10^{-13}$ Hz s$^{-1}$, $I_{45}$ the moment of inertia of the 
neutron star in units of $10^{45}$ g cm$^2$, and $m$ is the neutron 
star mass in units of $M_\odot$. This would correspond to a 
bolometric luminosity of $\sim 7 \times 
10^{37}$ erg s$^{-1}$, that is about an order of magnitude higher than 
the X--ray luminosity inferred from the observed X--ray flux, assuming 
a distance of 5 kpc. Once we will have a direct, independent, estimate 
of the distance to the source, we will have the possibility to test the
$\dot M$ vs.\ X--ray luminosity relation, torque theories and/or the 
physical parameters of the neutron star.

Other AMSPs show clear parabolic trend of the pulse phase delays as
a function of time during X--ray outburst, some of them showing spin-up
(e.g.\ XTE J1807--294, $\dot \nu = 2.5(7) \times 10^{-14}$ Hz s$^{-1}$, 
\cite{Riggio2008}; XTE J1751--305, $\dot \nu = 3.7(1.0) \times 10^{-13}$ 
Hz s$^{-1}$, \cite{Papitto2008}; IGR J17511--3057, $\dot \nu = 1.6(2) 
\times 10^{-13}$ Hz s$^{-1}$, \cite{Riggio2011}) while others showing 
spin-down (e.g.\ XTE J0929--314, $\dot \nu = -9.2(4) \times 10^{-14}$ Hz 
s$^{-1}$, \cite{Galloway2002}; XTE J1814-338, $\dot \nu = -6.7(7) \times 
10^{-14}$ Hz s$^{-1}$, \cite{Papitto2007}; IGR J17498--2921, $\dot \nu = 
-6.3(1.9) \times 10^{-14}$ Hz s$^{-1}$, \cite{Papitto2011b}). 
Those sources showing spin-down suggest the
possibility of an interaction of the neutron star magnetic field 
with the accretion disc outside the co-rotation radius (the radius in
the disc where the Keplerian frequency equals the neutron star spin
frequency). In fact the magnetic field lines can be threaded into the 
accretion disc and dragged by the high conductivity plasma so that an 
extra torque due to magnetic stresses has to be expected (see e.g.\ 
\cite{Wang1987}). In this case, an estimate of the magnetic field strength 
can be derived from the measured spin-down rate (see e.g.\ \cite{DiSalvo2007}). 
However, there is not a general consensus on the interpretation 
of these phase residuals. Another possibility is that these are due to
a timing noise caused by a pulse phase offset that varies in correlation 
with X--ray flux, such that noise in flux translates into timing noise 
\cite{Patruno2009b}. In this case, much stringent limits result on 
the spin derivatives in these sources (see \cite{Patruno2012a}, and 
references therein). Although for some sources clear correlations 
are observed between abrupt jumps in the pulse phases and sharp variations 
in the X--ray flux, it is not clear yet how much of the phase variations 
can be ascribed to an accretion-rate-dependent hot spot location. 

Certainly, the most debated case is SAX J1808.4-3658 whose phase variations
are strongly dominated by timing noise. The pulse phase delays show a very 
puzzling behaviour, since a rather fast phase shift, by approximately 0.2 
in phase, is present at day 14 from the beginning of the 2002 outburst
\cite{Burderi2006}. Interestingly, 
day 14 corresponds to a steepening of the exponential decay with time of 
the X--ray flux. However, analysing separately the phase delays of the 
fundamental and second harmonic of the pulse profile, 
\cite{Burderi2006} noted that the phase delays of the harmonic did 
not show any evidence of phase jumps. 
This is not an effect of the worse statistics of the phase delays derived 
from the harmonic, which of course show larger error bars. 
 
\begin{figure}[h!]
\sidecaption
\includegraphics[scale=.3]{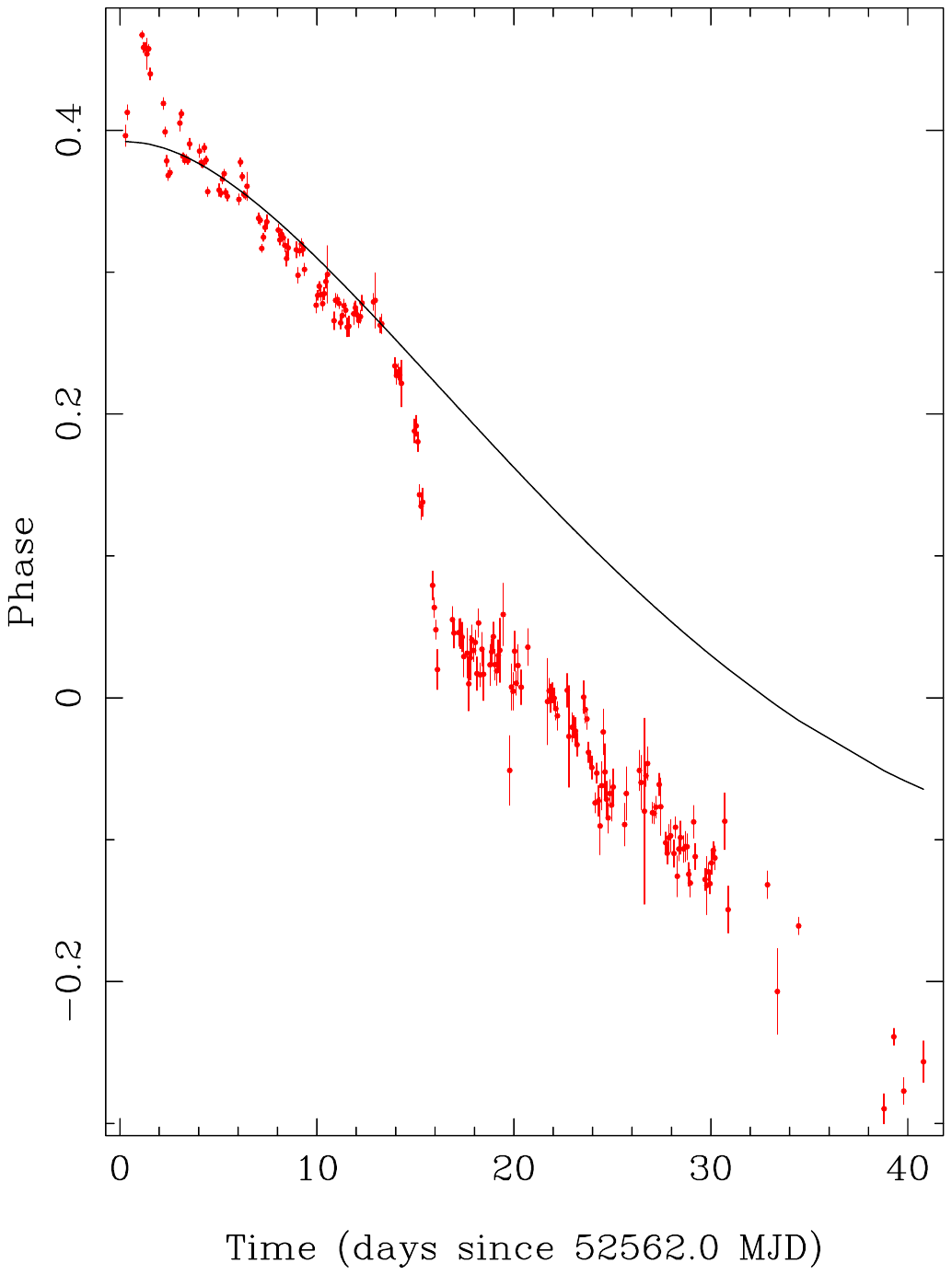}
\includegraphics[scale=.3]{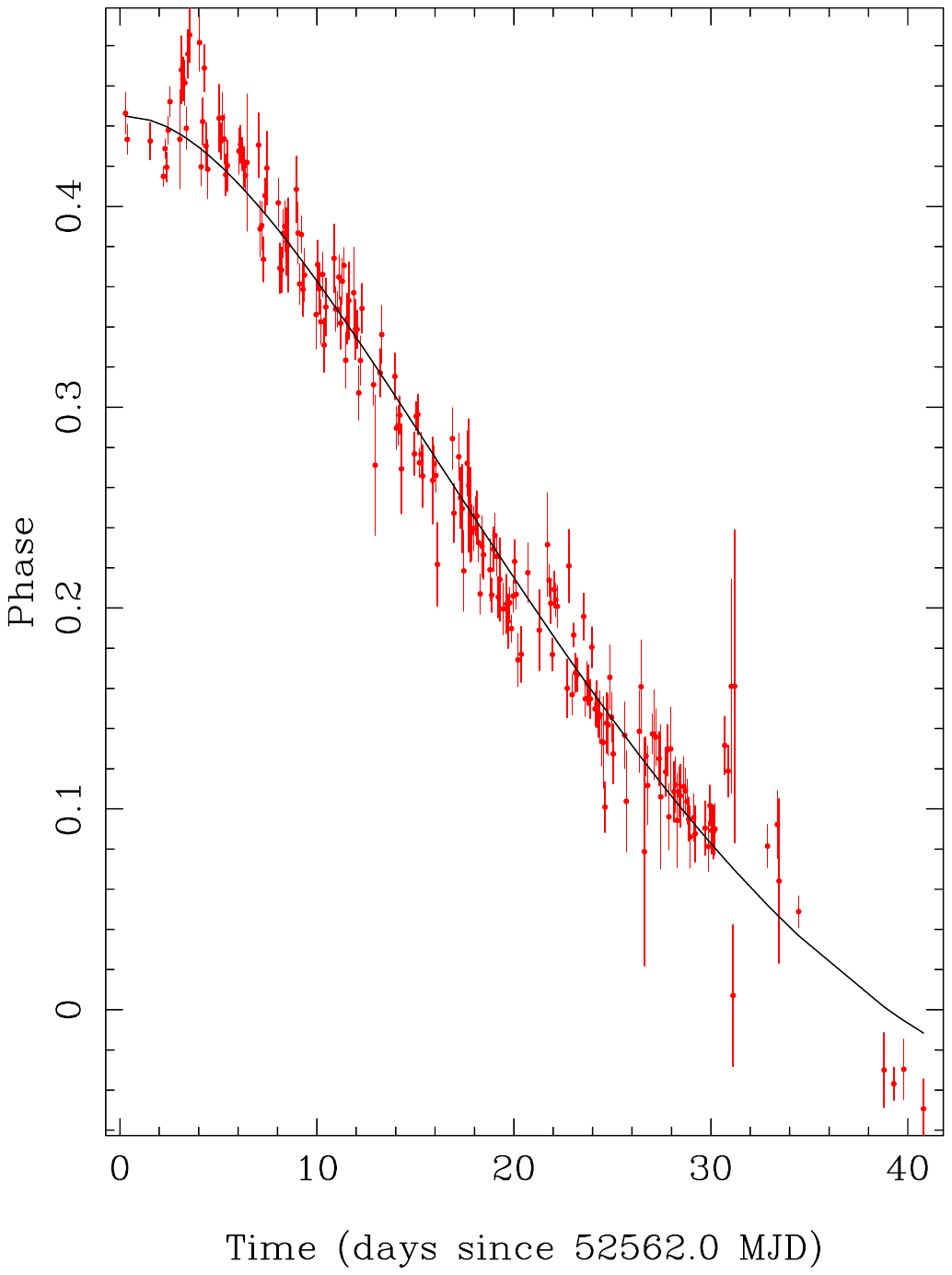}
\caption{{\bf Left:} Phase vs. time for the fundamental of the pulse frequency of 
SAX J1808.4-3658. {\bf Right:} Phase vs. time for the first harmonic of the pulse 
frequency of SAX J1808.4-3658. On top of the data, the best-fit function (including 
both the spin-up due to accretion and the spin-down at the end of the outburst) is 
plotted as a solid line (from \cite{Burderi2006}).}
\label{fig:1}       
\end{figure}

This means that the phase jump in the fundamental is not 
related to an intrinsic spin variation (which would have affected the 
whole pulse profile), but is instead caused by a change of the shape 
of the pulse profile (perhaps related to the mechanism causing 
the increase of the steepness of the exponential decay of the X--ray 
flux). 
On the other hand, from the fitting of the phase delays of the second harmonic, 
under the hypothesis that these are a better trace of the spin of the neutron 
star, \cite{Burderi2006} find that the source shows a spin-up 
at the beginning of the outburst with $\dot \nu_0 = 4.4(8) \times 10^{-13}$ 
Hz s$^{-1}$, corresponding to a mass accretion rate of $\dot M \sim 1.8 \times 10^{-9}\, 
M_\odot$ yr$^{-1}$, and a constant spin-down, with $\dot \nu_{sd} = 7.6(1.5) \times 
10^{-14}$ Hz s$^{-1}$, dominating the phase delays at the end of the outburst. 
In this case, the mass accretion rate inferred from timing is only a factor 
of 2 larger than the observed X--ray luminosity at the beginning of the 
outburst, that is $\sim 10^{37}$ erg s$^{-1}$.
The spin-down observed at the end of the outburst can be interpreted  
as due to a threading of the accretion disc by the neutron star magnetic 
field outside the co-rotation radius. Of course, in agreement with the
expectation, the threading effect appears to be more relevant at the end 
of the outburst, when the mass accretion rate significantly decreases. 
In this case the magnetic moment, $\mu$, of SAX J1808.4–3658 can be 
evaluated from the measured value of the spin-down, using the relation
(see \cite{Rappaport2004}):
\begin{equation}
\label{threading}
2\pi I \dot \nu_{sd} \equiv { \mu^2 \over 9 r_{\rm cor}^3}
\end{equation}
where $r_{\rm cor}$ is the co-rotation radius. The magnetic field found in 
this way is $B = (3.5 \pm 0.5) \times 10^8$ G, perfectly in agreement with other,
independent, constraints (e.g.\ \cite{Burderi2006}).

The fact that the second harmonic shows more regular phase residuals with
respect to the fundamental has also been observed in other AMSPs (e.g.\ 
\cite{Riggio2008,Riggio2011,Papitto2012}) and may indicate 
that the second harmonic is a good tracer of the neutron star spin frequency.
A simple model proposed by \cite{Riggio2011b} (see also \cite{Papitto2012}) 
may explain a similar behavior in terms of modest variations of the 
relative intensity received by the two polar caps on to the neutron star 
surface, in the hypothesis that the two spots emit a signal of similar 
amplitude and with a similar harmonic content. 
The sum of the two signals 
(the total profile) will be the result of a destructive interference 
for what concerns the fundamental frequency, since it is the sum 
of two signals with a phase difference of $\sim \pi$. A constructive 
interference develops instead for the second harmonic of the total 
profile, since it is the sum of two signals with the same phase. 
The destructive interference regarding the fundamental frequency leads 
to large swings of the phase of the fundamental of the total profile 
due to modest variations of the relative intensity of the signals 
emitted by the two caps. In this case swings up to 0.5 phase cycles 
can be shown by the phase computed over the fundamental frequency 
of the observed profile, without correspondingly large variations 
of the phase of the second harmonic. Interestingly, IGR J00291+5934,
showing a much more regular behaviour of the fundamental, shows a nearly 
sinusoidal pulse profile, with very little harmonic content (e.g. 
\cite{Galloway2005,Burderi2007}).

\subsection{Long-term spin evolution}
\label{spin}

For AMSPs for which more than one outburst has been observed, it is possible 
to derive their long term spin evolution comparing the averaged spin frequency 
measured in each outburst. To date only six AMSPs have been monitored 
with high time resolution instruments in different outbursts: SAX J1808.4-3658, 
IGR J00291+5934, XTE J1751--305, Swift J1756.9--2508, NGC6440 X-2 and SAX 
J1748.9--2021 (although, with relatively low S/N and short outburst duration 
in the latter two sources).
The best constrained is SAXJ1808.4-3658, for which secular spin evolution have 
now been measured over a 13 year baseline and shows a constant long-term
spin-down at a rate of $\sim -1 \times 10^{-15}$ Hz s$^{-1}$ (\cite{Hartman2008}, 
\cite{Hartman2009,Patruno2012c,Sanna2017b}). Because of the stability 
of the spin-down rate over the years, the most likely explanation appears to be loss 
of angular momentum via magnetic-dipole radiation, which is expected for a rapidly 
rotating neutron star with a magnetic field. The measured spin-down is consistent 
with a polar magnetic field of $(1.5 - 2.5) \times 10^{8}$ G. This is in agreement 
with the estimate above.

A spin down has also been measured for IGR J00291+5934 between the 2004
and 2008 outburst, at a rate of $-4.1(1.2) \times 10^{-15}$ Hz s$^{-1}$ 
(\cite{Patruno2010,Papitto2011c,Hartman2011}), larger 
than that observed in SAX J1808.4–3658, as expected given that IGR J00291+5934 
spins at a higher frequency. 
If interpreted in terms of magneto-dipole emission, the measured spin down 
translates into an estimate of the neutron star magnetic field of $(1.5-2) 
\times 10^8$ G. For the period between the 2008 and 2015 outbursts only an
upper limit to the frequency evolution could be derived, $|\dot \nu| \le 6
\times 10^{-15}$ Hz s$^{-1}$ ($3\sigma$ c.l., \cite{Sanna2017c}), compatible 
with the previous estimate.
Comparing the spin frequencies from 2002, 2005, 2007 and 2009 outbursts of
XTE J1751--305, \cite{Riggio2011c} report a spin down at a rate of 
$\sim (1.2) \times 10^{-15}$ Hz s$^{-1}$ and an inferred magnetic field of 
$\sim 4 \times 10^{8}$ G. Whereas for Swift J1756.9--2508 only an upper limit 
($|\dot \nu_{sd}| \le 2 \times 10^{-15}$ Hz s$^{-1}$), corresponding to a magnetic
field $\le 10^9$ G, has been reported \cite{Patruno2010b}.

The fact that the spin-down during quiescent periods is probably due to
magnetic-dipole radiation rises the interesting possibility that AMSPs may 
switch on as a radio pulsar during quiescence and ablate their donor. This 
is the so-called {\it hidden black widow} scenario proposed by \cite{DiSalvo2008} 
(see also \cite{Stella1994,Campana1998}) to explain the long-term orbital 
evolution (see next section).

\subsection{Orbital evolution}
\label{evolution}

The study of the orbital evolution in these systems is important to
constrain the evolutionary path leading to the formation of radio MSPs. 
These studies, however, require a large 
timespan of data in order to constraint the orbital period derivative.
Hence, the main difficulty is given by the fact that most of the AMSPs
rarely turn into X--ray outburst. For this reason, the best constraints 
on the orbital evolution in these systems come again from SAX J1808.4-3658,
which has shown seven X--ray outburst to date, allowing to follow its
orbital period over 17 years.

In the case of SAX J1808.4-3658 it is possible to see
a clear parabolic trend of the time of passage to the ascending node
versus time over the last 17 years (\cite{DiSalvo2008,Hartman2008, 
Burderi2009,Patruno2016,Sanna2017b}). Interpreting this 
parabolic term as the orbital period derivative, gives orbital expansion
at a quite high rate of $\dot P_{orb} = (3.4 - 3.9) \times 10^{-12}$ s s$^{-1}$
(see also \cite{Sanna2016} who report a marginally significant, strong 
orbital expansion in the AMSP SAX J1748.9--2021).
The observed orbital expansion implies a mass-radius index for the secondary 
$n < 1/3$ (see \cite{DiSalvo2008}). In the reasonable hypothesis that the 
secondary star is a fully convective star out of thermal equilibrium and 
responds adiabatically to the mass transfer, a mass-radius index of $n = 
-1/3$ can be assumed for the secondary. However, this derivative is a factor 
$\sim 70$ higher that the orbital derivative expected for conservative mass 
transfer, given the low averaged mass accretion rate onto the neutron star; 
since SAX J1808.4-3658 accretes for about 30 d every $2-4$ yr, the averaged 
X--ray luminosity from the source results to be $L_{obs} \sim 4 \times 10^{34}$ 
erg s$^{-1}$.

A non-conservative mass transfer can explain the large orbital period derivative 
if we assume a mass transfer rate of $\dot M \sim 10^{-9}\, M_\odot$ yr$^{-1}$, and 
that this matter is expelled from the system with the specific angular momentum 
at the inner Lagrangian point (see \cite{DiSalvo2008,Burderi2009}). 
In this case, the non-conservative mass transfer 
may be a consequence of the so-called {\it radio-ejection} model extensively  
discussed by \cite{Burderi2001}. The basic idea is that a fraction of the 
transferred matter in the disc could be swept out by radiative pressure of the 
pulsar. 
In this case, the fast spinning neutron star may ablate the companion during
quiescent periods (the so-called hidden black widow scenario proposed by 
\cite{DiSalvo2008}).
Alternatively, the large orbital period derivative observed in SAX J1808.4-3658 
can be interpreted as the effect of short-term angular momentum exchange between 
the mass donor and the orbit (\cite{Hartman2009,Patruno2012c}), resulting from 
variations in the spin of the companion star (holding the star out of synchronous 
rotation) caused by intense magnetic activity driven by the pulsar irradiation. 
This mechanism has been invoked by \cite{Applegate1994} (hereafter A\&S) 
to explain oscillating orbital residuals observed in some radio MSPs 
(see e.g.\ \cite{Arzoumanian1994}). In this case, the energy flow in the 
companion needed to power the orbital period change mechanism can be supplied 
by tidal dissipation. However, the A\&S mechanism envisages alternating epochs 
of orbital period increase and decrease, which is not yet observed from 
SAX J1808.4-3658. It also predicts that the system will evolve to longer 
orbital periods by mass and angular momentum loss on a timescale of $10^8$ yr
(for a 2-hr orbital period and a companion mass of $0.1-0.2\, M_\odot$),
and thus requires a strong orbital period derivative, similar to that inferred
from the quadratic trend observed in SAX J1808.4-3658. 
Therefore, also in the framework of the A\&S mechanism, most of the orbital 
period variation observed in SAX J1808.4-3658 is probably caused by loss 
of matter and angular momentum, i.e.\ by a non-conservative mass 
transfer (see \cite{{Sanna2017b}} for further discussions).
The next outbursts from this source will tell us whether the orbital period 
increase will turn into a decrease or, instead, the orbital expansion will 
continue, and this will be crucial in order to discriminate between the 
two possibilities sketched above.

Another puzzling result comes from IGR J0029+5934, which has orbital parameters
very similar to those of SAX J1808.4-3658, and is considered its {\it orbital twin}.
IGR J0029+5934 has shown only four outbursts since its discovery, but tight 
upper limits could be derived on its orbital period derivative, $|\dot P_{orb}|
< 5 \times 10^{-13}$ s s$^{-1}$ (90\% confidence level \cite{Patruno2017b}, see also
\cite{Sanna2017c}). 
This implies a much slower orbital evolution, on a timescale longer than 
$\sim 0.5$ Gyr, as compared to the fast orbital evolution of its twin, 
$\sim 70$ Myr.  
The orbital evolution observed in IGR J0029+5934 is compatible with the 
expected timescale of mass transfer driven by angular momentum loss via 
gravitational radiation, with no need of A\&S mechanism or non-conservative
mass transfer. In this case, it would be interesting to constrain the sign
of the orbital period derivative in order to get information on the mass-radius 
index of the donor star and to infer whether it is in thermal equilibrium 
(implying orbital contraction) or not (implying orbital expansion).

\subsection{Searches at other wavelengths}
\label{multiwave}

AMSPs are transient systems with inferred variations in the mass accretion 
rate onto the central source by a factor of $\sim 10^5$ between outbursts and 
quiescence. Since the magnetospheric radius expands as the mass accretion 
rate decreases, it is easy to see that, while during an X--ray outburst the 
magnetospheric radius is expected to be very close to the neutron star surface, 
during X--ray quiescence the magnetospheric radius may expand beyond the 
light-cylinder radius (where an object co-rotating with the neutron star attains the 
speed of light). In this case, it is expected that any residual accretion 
is inhibited by the radiation pressure and, consequently, it is plausible 
to expect that the neutron star turns-on as a radio MSP until a new outburst 
episode pushes the magnetospheric radius back again, quenching radio emission 
and initiating a new accretion phase. 
The compelling possibility that these systems could swiftly switch from 
accretion-powered to rotation-powered magneto-dipole emitters during quiescence 
gives the opportunity to study a phase that could shed new light on the not 
yet cleared up radio pulsar emission mechanism. However, such a behaviour has
been observed only recently, in the so-called transitional MSPs (see next 
sections), and in particular in the unique source IGR J18245--2452 \cite{Papitto2013a}, 
the long-sought-for ``missing link" between LMXBs and 
radio MSPs; a binary system containing a neutron star alternating accretion-powered 
phases, in which it behaves like an AMSP, to rotation-powered phases, in which 
it behaves like a radio MSP. 
However, despite the huge observational effort 
made to catch, in a transient LMXB, the transition between the accretion-powered 
regime and the rotation-powered regime during quiescence, no pulsed radio emission
has been found in other AMSPs.
The most embarrassing problem is certainly the lack of pulsed radio emission 
from these systems during quiescence: many transient LMXBs and AMSPs have been 
thoroughly searched in radio during quiescence with disappointing negative results 
(\cite{Burgay2003,Iacolina2009,Iacolina2010,Patruno2017}).  

\cite{Burderi2001} (see also \cite{Burderi2002}) have proposed a model, 
that naturally explains this non-detection, assuming that the radio pulsar 
mechanism switches on when a temporary significant reduction of the mass-transfer 
rate occurs. In some cases, even if the original mass transfer rate is restored, 
the accretion of matter onto the neutron star can be inhibited by 
the radiation pressure from the radio pulsar, which may be capable of ejecting 
out of the system most of the matter overflowing from the companion-star Roche 
lobe. In particular, \cite{Burderi2001} showed that in ``wide systems", i.e. 
systems with orbital periods longer than a few hours, and with a sufficiently 
fast spinning neutron star, the switch-on of the radio pulsar can prevent any 
further accretion, even if the original mass transfer rate is restored. On the 
other hand, ``compact'' systems (orbital periods below a few hours) should show a 
cyclic behaviour since, once the temporary reduction of the accretion rate ends, 
the radiation pressure of the pulsar is unable to keep the matter outside the 
light cylinder radius and accretion resumes. 

One of the strongest predictions of this model is the presence, during the 
radio-ejection phase, of a strong wind of matter emanating from the system: 
the mass released by the companion star swept away by the radiation pressure 
of the pulsar. This matter could cause strong free-free absorption in the radio 
band, hampering the detection of pulsed signals. A possible solution is then 
to observe at high radio frequencies (above 5--6 GHz, see \cite{Campana1998},
\cite{DiSalvo2003}), thus reducing the cross-section of free-free absorption 
which depends on $\nu^{-2}$. Note that the pulsed radio flux also decreases with 
increasing frequency, although it is easy to see that this decrease is less steep 
with respect to the decrease of the effects caused by free-free absorption
(see e.g.\ \cite{Burderi1994}).
This may explain why some AMSPs indeed have radio counterparts
(see \cite{Patruno2012a} for a review), although not pulsating. Interestingly,
\cite{Iacolina2010} report a peak at $4 \sigma$ significance for XTE J1751--305, 
obtained folding a radio observation of this source performed at Parkes 
radio telescope. This peak has a $40\%$ probability of not being randomly 
generated over the 40,755 trial foldings of the dataset corresponding to 
one of the two observations performed at 8.5 GHz. This result is not confirmed 
in the other observation (at the same frequency) 
and thus deserves additional investigation in the future.  
Alternatively, pulsating radio emission should be searched in systems 
with long orbital periods, in which the matter transferred by the companion 
star is spread over a wider orbit. 

Strong (indirect) evidences that a rotating magneto-dipole powers the quiescent 
emission of AMSPs, comes from observations of the quiescent emission from their 
identified optical counterpart. In the case of SAX J1808.4-3658, measures in the 
optical band show an unexpectedly large optical luminosity \cite{Homer2001}, 
inconsistent with both intrinsic emission from the companion star and 
X--ray reprocessing.  
The most probable explanation for 
the over-luminous optical counterpart of SAX J1808.4--3658 in quiescence, proposed 
by \cite{Burderi2003} and \cite{Campana2004}, is that the magnetic 
dipole rotator is active during quiescence and its bolometric luminosity, 
given by the Larmor's formula (eq.\ \ref{Larmor}), powers the reprocessed 
optical emission. Indeed, the optical luminosity and colours predicted by 
this model are perfectly in agreement with the observed values.

Similar results have been obtained for other AMSPs for which optical 
observations in quiescence have been performed. For XTE J0929--314 and
XTE J1814-338, optical photometry in quiescence showed that the donor was 
irradiated by a source emitting in excess of the X--ray quiescent luminosity 
of these sources, requiring an energy source compatible with the spin-down 
luminosity of a MSP \cite{DAvanzo2009}. IGR J00291+5934 showed 
evidences for a strongly irradiated companion in quiescence too \cite{DAvanzo2007}.

The precise spin and orbital ephemerides of AMSPs are of fundamental importance 
to allow deep searches of their counterparts in the gamma--ray band, which has
the advantage of not suffering the free-free absorption as in the radio band,
but the disadvantage of the paucity of photons, which requires folding over 
years in order to reach the statistics needed for detecting a pulsed signal.
Indeed, AMSPs, in analogy with MSPs (such as the so-called Black Widows and 
Red-Backs, detected in radio and most of them also in the gamma band; see 
\cite{Roberts2013} as a review) are expected to show coherent pulsations in the 
gamma band during X--ray quiescence. The detection of a possible gamma--ray 
counterpart of the AMSP SAX J1808.4-3658 from 6 years of data from the 
Fermi/Large Area Telescope has been recently reported \cite{deOnaWilhelmi2016}. 
The authors also searched for modulations of the flux at 
the known spin frequency or orbital period of the pulsar, but, taking into 
account all the trials, the modulation was not significant, preventing a 
firm identification via time variability. We expect that this result may be 
improved increasing the time span of the gamma--ray data.

\section{The missing link}
\label{sec:3}

As we described in the previous sections a clear path has been delineated bringing an old, slowly spinning neutron star 
to become a millisecond radio pulsar. The last link of the chain was supposed to be disclosed with the 
discovery of the (transient) accreting millisecond X--ray pulsar SAX J1808.4-3658 \cite{Wijnands1998}.
But this source, as well as all the other of this class, do not show a radio pulsar soul during their quiescence
(even if indirect hints for a turn on of the relativistic pulsar wind have been gathered, \cite{Burderi2003,Campana2004}).

A new way was paved by radio surveys. The Faint Images of the Radio Sky at Twenty-cm (FIRST) radio survey carried out at 
the Very Large Array (VLA) covered $\sim 10,000$ square degrees discovering nearly one million of radio sources. 
Matching radio sources with optical surveys Bond et al. \cite{Bond2002} reported the discovery of a new magnetic cataclysmic variable 
with radio emission: FIRST J102347.6+003841 (hereafter J1023). 
Optical spectroscopic studies revealed the presence of an accretion disc in 2001 through the presence of double-horned emission lines 
\cite{Szkody2003}, which led Thorstensen \& Armstrong \cite{Thorstensen2005} to suggest that J1023 could be a neutron star low-mass X--ray binary.
A bright millisecond radio pulsar (1.69 ms) was discovered in 2007 coincident with J1023 \cite{Archibald2009}. 
No signs of an accretion disc are discernible any more but the pulsed radio signal was eclipsed sporadically along the 4.8 hr orbit.
This was the first system testifying for the alternating presence of an accretion disc and of a millisecond radio pulsar, thus 
providing the first indirect evidence that the two souls can live within the same object.

Nature did even better. IGR J18245--2452 (J18245 hereafter) was discovered as a new X--ray transient in the globular cluster M28.
XMM-Newton pointed observations revealed a pulsed signal at 3.93 ms making of J18245 an accretion X--ray pulsar, 
modulated at 11 hr orbital period \cite{Papitto2013a}. Coincidentally, a radio pulsar was detected during a radio survey of M28 with 
the same spin period and binary orbital period (PSR 18245--2452I). 
J18245 closed definitely the chain: it is the first pulsar ever detected in X--rays and at radio wavelengths. 
Twenty days after the  end of the outburst J18245 was detected again as a radio pulsar, closing the loop and demonstrating that 
the transition between the rotation-powered and the accretion-powered regime can occur on short timescales.

\begin{figure}[h!]
\begin{center}
\includegraphics[scale=0.6]{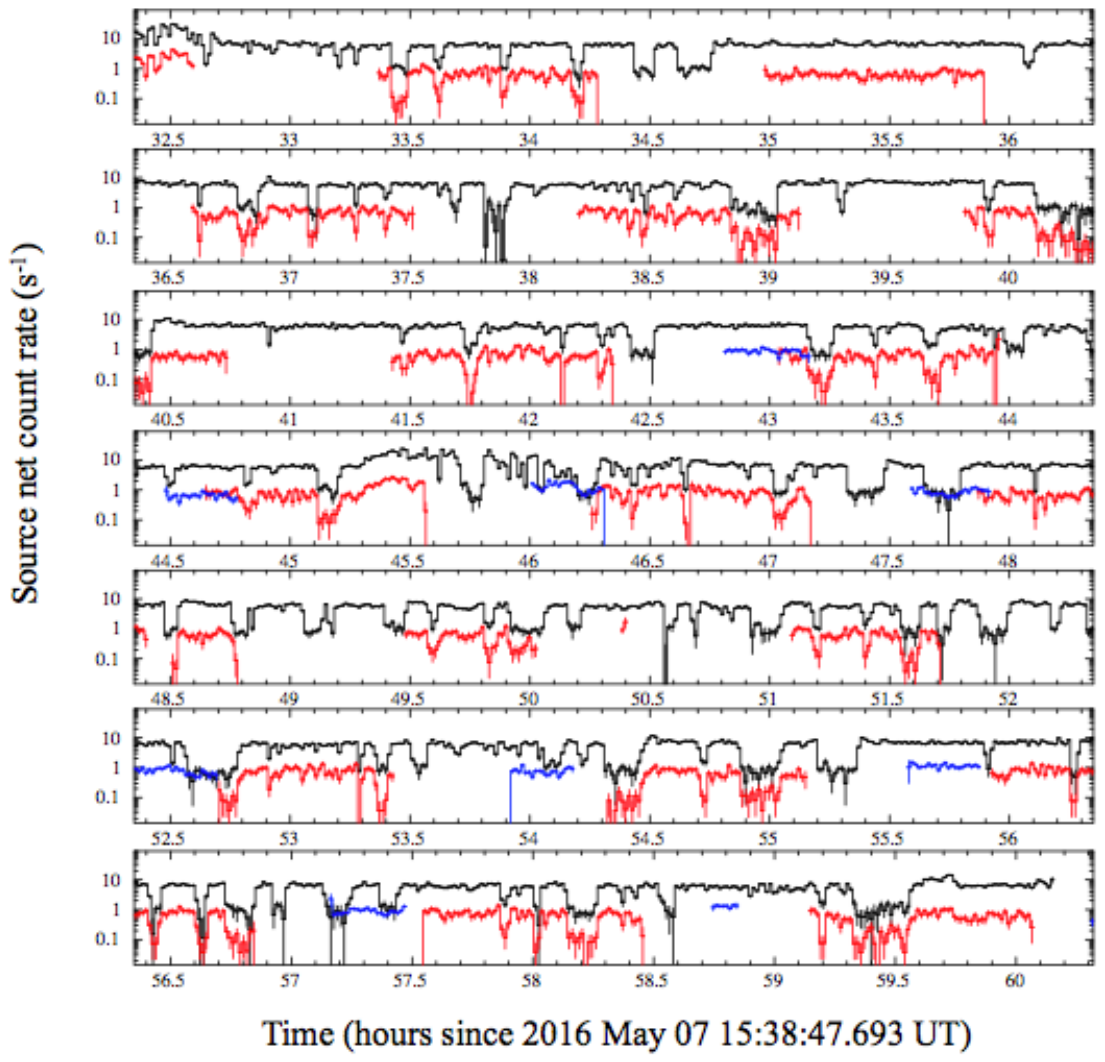}
\end{center}
\vskip -3truecm
\caption{Background-subtracted and exposure-corrected light curves of
J1023 obtained with the XMM-Newton EPIC cameras (0.3--10 keV; black data), NuSTAR FPMA + FPMB (3--79 keV; red data) 
and Swift UVOT (UVM2 filter; blue data) during the time interval covered by XMM-Newton. For plotting purpose, light curves are shown 
with a binning time of 50 s and the vertical axis is plotted in logarithmic scale (from \cite{CotiZelati2018}).}
\label{xcurve}       
\end{figure}

Incidentally, in mid 2013 the radio monitoring of J1023 failed to detect the radio pulsar any more.
Simultaneously the optical, X--ray and gamma--ray flux increased by a factor of $\sim 70$ and $\sim 5$, respectively,  
even if the source did not enter in a proper outburst state \cite{Patruno2014,Stappers2014}. 
During this active state J1023 displayed a very peculiar behaviour with three different states:
a high state during which X--ray pulsations are detected, a low state (a factor of $\sim 7$ dimmer) during which 
no pulsations were detected (and the upper limit on the pulsed fraction is lower than what observed during the high state) 
and a flaring state (with no pulsations too), outshining both states during which the source wildly varies. 
Transitions occur on a $\sim 10$ s timescale in X--rays 
\cite{Archibald2015,Bogdanov2015}. More details will be provided in the next section (see also Figure \ref{xcurve}). 
Another source distinctly showed this behaviour was XSS  J12270--4859 \cite{Papitto2015} (J12270 in the following), so that 
this puzzling X--ray light curve has become the archetypal way to identify a ``transitional" X--ray pulsar. 

\section{Madamina, il catalogo \`e questo (``Don Giovanni'', Mozart)}
\label{sec:4}

There are 4 transitional pulsars and a few candidates. At the moment of writing (October 2017) J1023 is in an active state. 
J18245 and J12270 are in quiescence, shining as radio pulsars. We summarise their main characteristics in Table \ref{table_tmps}.
 
\subsection{PSR J1023+0038}
\label{J1023}

J1023 was discovered as a peculiar magnetic cataclysmic variable with radio emission \cite{Bond2002}. 
Optical studies revealed signs for the presence of an accretion disc through double horned emission lines in 2001 \cite{Szkody2003,Wang2009}.
Thorstensen \& Armstrong \cite{Thorstensen2005} performed photometric and spectroscopic optical observations campaign. 
The spectrum showed mid-G star features and the disappearance of the accretion disc signatures.  Photometry showed a 
smooth orbital modulation at 4.75 hr, with colour changes consistent with irradiation (but with no emission lines). 
Radial velocity studies and modelling of the light curves led them to show that the primary should be more massive than the Chandrasekhar mass, 
thus pointing to a LMXB, rather than a cataclysmic variable. Homer et al. \cite{Homer2006} used XMM-Newton data, lack of optical circular polarisation, 
and optical spectroscopic data to confirm this picture. 
 
This changed with the discovery of a radio pulsar coincident with J1023. Archibald et al. \cite{Archibald2009} discovered a bright, fast spinning (1.69 ms) MSP 
during a low-frequency pulsar survey carried out with Green Bank Telescope in 2007. The pulsar is eclipsed during a large fraction of the 
orbital period (at orbital phases $0.10-0.35$). Due to its proximity and radio brightness it was possible to derive a very precise distance 
of  $1.37\pm 0.04$ kpc based on its radio parallax \cite{Deller2012}. XMM-Newton observations then revealed a $\sim 9\times 10^{31}$ erg 
s$^{-1}$ (0.5--10 keV) source showing pulsed emission at the radio period. The root-mean-squared pulsed fraction in the 0.25--2.5 keV energy range 
is $11\pm2\%$, whereas a $3\,\sigma$ upper limit of $20\%$ is obtained at higher energies \cite{Archibald2010,Bogdanov2011}.
The neutron starÕs parameters were determined thanks to the radio pulsar signal, with a spin period of 1.69 ms and a dipolar magnetic 
field of $9.7\times10^7$ G \cite{Archibald2009,Deller2012}.

\begin{figure}[h!]
\begin{center}
\includegraphics[scale=10.]{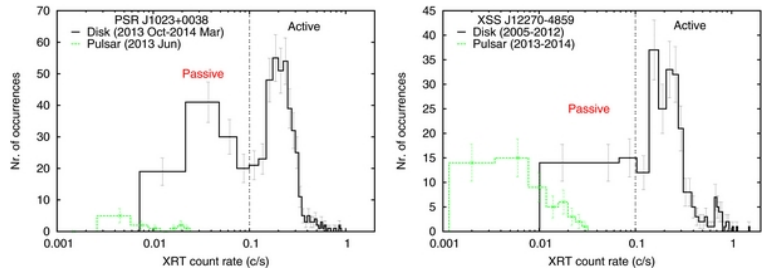}
\end{center}
\caption{Count rate distribution in the 0.3--10 keV XRT light curves of J1023 (left) and  and J12270 (right).
The disc state (solid black histogram) and pulsar state (dashed green histogram) are 
shown separately on the same scale. The horizontal dotted line at 0.1 c s$^{-1}$ for both sources 
marks the boundary between disc-active and disc-passive states (labeled in black and red, respectively,
from \cite{Linares2014}).}
\label{bimodal}       
\end{figure}

Surprisingly, J1023 started not being detected in the radio band around June 2013 \cite{Patruno2014,Stappers2014}. 
Contemporaneously, the weak 
$\gamma$-ray flux detected by Fermi increased by a factor of $\sim 5$ \cite{Stappers2014,Takata2014,Torres2017}. 
This new state persisted in time and J1023 is still in this active state now (October 2017) with no signs of change.
Together with the disappearance of the radio signal, J1023 brightened at all the other wavelengths. In the X--rays (0.5--10 keV) it brightened 
by a factor of $\sim 30$, with some flaring activity reaching $10^{34}$ erg s$^{-1}$ \cite{Patruno2014,Takata2014,CotiZelati2014,Bogdanov2015}. 
In the optical, J1023 brightened too by $\sim 1$ mag and showed again the presence of several broad, double-horned
emission lines typical of an accretion disc \cite{Halpern2013,CotiZelati2014}.

At variance with any other LMXBs, J1023 shows a highly variable active state. A simple histogram of the observed count rates shows a 
bimodal distribution \cite{Linares2014} (see Figure \ref{bimodal}). 
A closer look (thanks to XMM-Newton observations) revealed the existence of three distinct states 
\cite{Bogdanov2015,Archibald2015}. These can be characterised as: 
\begin{itemize}
\item a high state with a 0.3--80 keV luminosity $L_X \sim 7 \times 10^{33}$ erg s$^{-1}$ \cite{Tendulkar2014} occurring for $\sim 80\%$ of the 
time and during which X--ray pulsations at the neutron star spin period are detected with a r.m.s. pulsed fraction of $8.1\pm0.2\%$ 
(0.3--10 keV) \cite{Archibald2015};
\item a low state with a 0.3--80 keV luminosity $L_X \sim 10^{33}$ erg s$^{-1}$ occurring for $\sim 20\%$ of the time and during which 
pulsations are not detected with a $95\%$ r.m.s. upper limit $\lsim 2.4\%$ (0.3--10 keV), much smaller than the detection during the high state;  
\item a flaring state during which sporadic bright flares occur reaching luminosities as high as  of  $\sim 10^{35}$ erg s$^{-1}$, 
with no pulsation too. 
\end{itemize}

The transition between the high and the low states is very rapid, on a $\sim 10$ s timescale and looks symmetric (ingress time equals 
to the egress time). It is not clear if similar variability has also been detected in the optical band, superimposed to the orbital modulation \cite{Shahbaz2015,Jaodand2016}. 
A phase-connected timing solution shows that the neutron star is spinning down at a rate $\sim 30\%$ faster than the spin down due to rotational 
energy losses \cite{Jaodand2016}.

The 0.1--300 GeV flux of increased by a factor of $\sim 5$ after the transition to the active state with a steep power law photon index of $2.5-3$ 
\cite{Stappers2014,Takata2014,Deller2015}. 
Above 300 GeV VERITAS put instead an $95\%$ upper limit of $7\times 10^{-13}$ erg cm$^{-2}$ s$^{-1}$ on the flux \cite{Aliu2016}.

Radio monitoring observations during the active state revealed a rapidly variable, flat spectrum persistent source. This emission is likely
suggesting synchrotron emission as the origin. If this is in form of a jet or more generally of a propeller outflow is unknown \cite{Deller2015}.
In addition, based on existing correlation in neutron star LMXBs between the radio luminosity and the X--ray luminosity, J1023 
is brighter in radio, possibly suggesting a less efficient X--ray emission \cite{Deller2015}.
Baglio et al. \cite{Baglio2016} measured a linear polarisation of $0.90\pm0.17\%$ in the $R$ band. In addition, the phase-resolved 
$R$-band curve shows a hint for a sinusoidal modulation. Lacking the Spectral Energy Distribution a red/nIR excess (characteristic of jet emission), 
the polarised emission likely comes from Thomson scattering with electrons in the disc.

Simultaneous X--ray (Chandra) and radio (VLA) monitoring showed a strong anti-correlated variability pattern, with radio emission strongly rising during 
X--ray low states \cite{Bogdanov2017}. A more articulated observing campaign involving XMM-Newton, NuSTAR, and Swift showed that X--rays al soft 
and hard (up to 80 keV) are strongly corrected with no lag, whereas X--rays and UV are not correlated \cite{CotiZelati2018}.

Surprisingly, Ambrosino, Papitto et al. \cite{Ambrosino2017} discovered optical pulsations in J1023 during the active X--ray state.
The pulsed fraction is at a level of $\sim 1\%$. Optical pulsation is present only in the high mode as X--ray pulsations (A. Papitto, private communication). 
This optical pulsed emission is puzzling. Ambrosino, Papitto et al. 
\cite{Ambrosino2017} convincingly showed that this emission cannot be explained by the cyclotron mechanism and cautiously favour a 
rotation-powered regime mechanism.

\begin{table}
\caption{Parameters of transitional millisecond pulsars.}
\label{tab:1}       
%
%
\begin{tabular}{ccccccccc}
\hline\noalign{\smallskip}
Source & Spin Period&Orbital period&Distance &Companion       & DM  & Fermi      & X--ray          & Radio \\
              & (ms)              & (hr)                 & (kpc) & mass ($\msole$)& (pc cm$^{-3}$)            & detection& pulsations & pulsations\\
\noalign{\smallskip}\svhline\noalign{\smallskip}
J1023   & 1.69              &  4.75              & 1.37        & $\sim 0.24$ & 14.3  &  Y             &   Y                & Y\\
J12270 & 1.69              &  6.91             & 1.4           & $\sim 0.25$ &  43.4 & Y              &  Y                & Y\\
J18245 & 3.93              & 11.03            &5.5 (M28)  & $\sim 0.2$   &   119       & N             & Y                 & Y \\
 J154439& --                 & $\sim 5.3$    &   --           &    --               &   --     & Y             & N                 & N \\
\noalign{\smallskip}\hline\noalign{\smallskip}
\end{tabular}
\label{table_tmps}
\end{table}

\subsection{IGR J18245--2452}
\label{IGR}

IGR J18245--2452 (J18245 in the following) was discovered by INTEGRAL/ISGRI during observations of the Galactic centre region \cite{Eckert2013}.
J18245 lies in the globular cluster M28 at a distance of $\sim 5.5$ kpc \cite{Heinke2013,Romano2013,Homan2013}. 
At this distance the peak outburst luminosity is $\sim 10^{37}$ erg s$^{-1}$ (0.5--100 keV, e.g. \cite{DeFalco2017}). 
This luminosity led J18245 to be classified as a classical X--ray transient (i.e. not faint). 
A thermonuclear (type I) X--ray burst from J18245 was detected by Swift/XRT \cite{Papitto2013b,Linares2013}. This marked the presence
of a neutron star in the system. Further type I bursts were observed during the same outburst by MAXI \cite{Serino2013} and INTEGRAL 
\cite{DeFalco2017}. 
During an XMM-Newton observation, Papitto et al. \cite{Papitto2013a} discovered a coherent periodicity in the X--ray flux at 3.9 ms. 
The pulsed signal is also modulated through Doppler shifts at the binary orbital period of 11.0 hr, induced by a $\sim 0.2\msole$ 
companion star. Papitto et al. \cite{Papitto2013a} were also able to associate the X--ray pulsar to a known radio pulsar previously discovered in M28, 
PSR J1824--2452I \cite{Manchester2005}, with the same spin and orbital periods. This provides the first direct evidence for a switch 
between an accretion-powered neutron star and a rotation-powered radio pulsar.  The reactivation of the radio pulsar was very fast, with 
the detection of a pulsed signal less than two weeks after the end of the outburst.

\begin{figure}[h!]
\begin{center}
\includegraphics[scale=0.5]{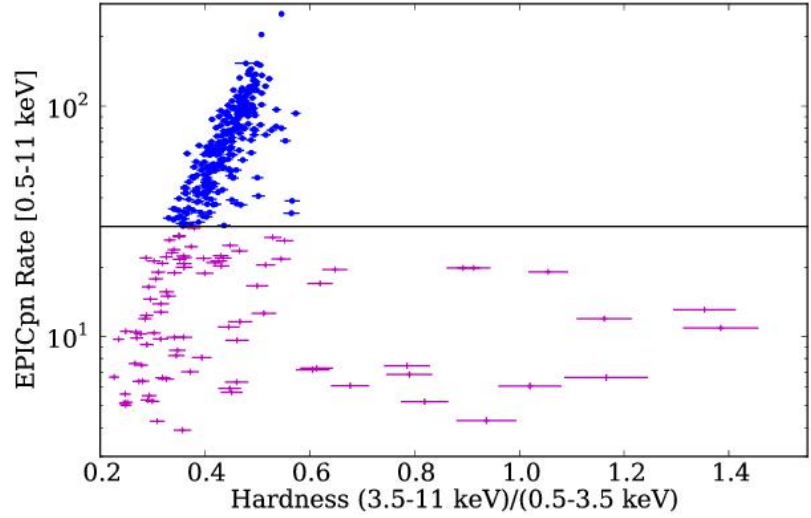}
\end{center}
\caption{Hardness-intensity diagram built by using two XMM-Newton observations  with time bin of 200 s. The black solid line separates the 
different intensity states: the points represented in magenta and blue have a count rate lower and higher than 30 c s$^{-1}$, respectively 
(from \cite{Ferrigno2014}).}
\label{18245branch}       
\end{figure}

A very peculiar state was observed when the source luminosity reached a mean level of a few $10^{36}$ erg s$^{-1}$.
Rapid variation by a factor up to $\sim 100$ were observed during two XMM-Newton observations \cite{Ferrigno2014}.
In a hardness-intensity diagram two branches can be identified (see Figure \ref{18245branch}; \cite{Ferrigno2014}).
The brighter branch (blue branch) showed a tight correlation between hardness and intensity (as well as pulsed fraction).
Below a threshold of $\sim 30$ c s$^{-1}$ in the pn instrument in addition to this branch a new one (magenta branch), 
appeared with scattered points at higher hardness. J18245 varied by a factor of $\sim 100$ on a timescale as short as a few seconds. 
A spectral analysis at different spectral hardness shows a clear decrease in the power law spectral index 
from $\Gamma\sim 1.7$ to $\Gamma\sim 0.9$ and the disappearance of a black body component.  The hardest spectrum is 
better described by partially covering power law model and it is achieved only occasionally at low count rates.
A pulsed signal is always detected and the pulsed fraction is tightly correlated with the source count rate. 
In the magenta branch there is however no correlation of the hardness with the pulsed fraction, which is always at a level of $5-10\%$
\cite{Ferrigno2014}.

M28 has been observed several times, however only thanks to the spatial resolution of the Chandra optics it is possible to 
gain information on the quiescence of J18245. 
Chandra observed M28 in 2002 finding J18245 in a quiescent state. The X--ray spectrum is well described by a simple 
power law model with a hard photon index $\Gamma\sim 1.2$. The 0.5--10 keV unabsorbed luminosity is $\sim 2\times 10^{32}$ 
erg s$^{-1}$ \cite{Linaresetal2014}. No room for a soft component is left with an upper limit on the 0.1--10 keV luminosity of 
$\lsim 7\times 10^{31}$ erg s$^{-1}$. 
In a way similar to J1023 (but before J1023), also J18245 shows two different low-luminosity active states. 
They are readily apparent in a long Chandra observation taken in 2008 \cite{Linaresetal2014}.
The high and low state 0.5--10 keV luminosities are $\sim 4\times10^{33}$ erg s$^{-1}$ and $\sim 6\times 10^{32}$  erg s$^{-1}$, 
respectively, with a factor of $\sim 7$ luminosity change. 
The spectra in the two states are fully compatible with a power law with photon index $\Gamma\sim 1.5$.
Given the lower Chandra count rate, mode switching has been measure to occur on a timescale of $\lsim 200$ s.

The optical counterpart has been identified thanks to HST images \cite{Pallanca2013}. 
The companion star has been detected during both quiescence and outburst, showing a two magnitude increase and 
the presence of the H$\alpha$ line, indicating that accretion is taking place.

PSR J1824Ð2452I is known as a radio pulsar in M28 (detected during the quiescent period), but its observations were 
only sporadic and with large eclipses, variable from orbit 
to another as often happens in redbacks. In addition, the acceleration induced by the motion in the globular cluster prevents us from a firm 
measurement of the magnetic field.

\subsection{XSS  J12270--4859}
\label{XSS}

XSS  J12270--4859 (J12270 in the following) was discovered by the Rossi X--ray Timing Explorer during the high latitude slew survey 
\cite{Sazonov2004}. Based on the presence of optical emission lines J12270 was initially classified 
as a cataclysmic variable hosting a magnetic white dwarf \cite{Masetti2006,Butters2008}, as for J1023.
Unusual dipping and flaring behaviour led several authors to suggest a different classification involving a neutron star in a LMXB 
\cite{Pretorius2009,Saitou2009,deMartino2010,deMartino2013,Hill2011,Papitto2015}. 
de Martino et al. \cite{deMartino2010} (see also \cite{Hill2011}) were the first to associate J12270 to a relatively bright gamma--ray source detected 
by Fermi-LAT and emitting up to 10 GeV (3FGL J1227.9--4854).
The source was puzzling and has been observed during this active state by XMM-Newton. An erratic behaviour typical of 
TMSP has been observed \cite{deMartino2010}. Despite flaring and dips,  J12270 was stable at least over a 7 yr period 
\cite{deMartino2013}. Dips correspond to the low state of J1023.
A detailed analysis of the dips in J12270 disclosed three different types of dips: soft dips, dips with no spectral changes with respect to the 
active state and hard dips following flares.
The pn spectrum can be modelled with a simple power law resulting in spectral indexes of: $1.64 \pm 0.01$ (active state); $1.65\pm0.03$ (flares);
$1.71\pm0.04$ (dips), and $0.74\pm0.08$ (post-flare dips; \cite{deMartino2013}).
The 0.2--100 keV mean luminosity is $10^{34}$ erg s$^{-1}$ at a distance of 1.4 kpc. 
The ratio between the active and dip 0.2--10 keV luminosity is $\sim 5$,
with the active 0.2--10 keV luminosity being $4\times 10^{33}$ erg s$^{-1}$ \cite{deMartino2013}.

The XMM-Newton Optical Monitor (OM) was operated in fast timing mode ($U$ and $UVM2$ filters) allowing for a strict comparison with X--ray data. 
The mean $U$ magnitude was 16.6. Orbital modulation is readily apparent in the OM data. On top of this dips and flares are also present 
in the OM data, with a drop in the UV count rate by a factor of $\sim 1.4$ \cite{deMartino2013}.
A lag analysis was also possible, showing no lag among soft (0.3--2 keV) and hard (2--10 keV) X--ray flux. 
The cross-correlation between $U$-band and X--rays show no lag, but indications that flares last longer in the optical-UV bands.
A cross-correlation analysis on selected dips shows that UV and X--ray dips occur almost simultaneously, but the shape of UV dips is 
shallower, with a smoother decay and rise \cite{deMartino2013}.
The optical spectrum shows prominent H$\alpha$, H$\beta$, He I, He II, and Bowen-blend N III/C III  lines, as well as signs of irradiation 
\cite{deMartino2014}.

As J1023 during its active state also J12270 has been detected by Fermi in the 0.1--300 GeV, with a 0.1--10 GeV flux of $4\times10^{-11}$
erg cm$^{-2}$ s$^{-1}$ and a steep power law spectrum with $\Gamma\sim 2.2$ and a cut-off energy of 8 GeV \cite{Johnson2015,deMartino2010}. 
Faint non-thermal radio emission was also detected with a flat spectral index \cite{Masetti2006,Hill2011}.

J12270 remained stable at all wavelengths for about a decade up to 2012 November/December, when a decline in flux at all bands 
was reported \cite{Bassa2014,Bogdanov2014}. 
This is in all respects similar to the transition of J1023 in the opposite sense: J12770 changed from an active state to a quiescent state, 
whereas the opposite transition was observed in J1023 in June 2013.
J12270 decreased its optical brightness by 2 magnitudes, with all the optical emission lines disappearing. 
The X--ray spectrum hardened considerably with the spectrum described by power law with a photon index of $\Gamma=1.1$ and a
0.3--10 keV unabsorbed luminosity of $\sim 2\times 10^{32}$ erg s$^{-1}$, resulting in a 
factor of $\sim 20$ decrease with respect to the high state and $\sim 4$ with respect to the low state \cite{deMartino2015,Bogdanov2014}.
At GeV energies the 0.1--100 GeV flux decreased by a factor of $\sim 2$ and the spectrum hardened to a power law with photon index 
1.7 with a cut-off at 3 GeV \cite{Johnson2015}.

Radio observations revealed the presence of a millisecond radio pulsar at the position of J12270. The neutron star is spinning at 1.69 ms
and has a magnetic field of $1.4\times 10^8$ G (for a rotational energy of $9\times 10^{34}$ erg s$^{-1}$; \cite{Roy2015}). The radio 
signal is absorbed for a large fraction of the orbit. After this discovery pulsations were searched at other frequencies and in older data.
Pulsed emission was observed at GeV energies with a single peak emission nearly aligned with the radio main peak \cite{Johnson2015}.
The pulsed signal was not detected in the X--ray band with a $3\,\sigma$ upper limit on the rms pulsed fraction of $\sim 7\%$ (full band) or 
$\sim 10\%$ (0.5--2.5 keV; \cite{Papitto2015}). Papitto et al. \cite{Papitto2015} reanalysed XMM-Newton data during the high state and, knowing the pulse period, 
successfully detected a coherent pulsation.  Pulsations were detected at an rms amplitude of $\sim 8\%$, with a second 
harmonic stronger than the fundamental frequency. The amplitude is similar in the soft and hard X--ray bands.
Polarimetric optical observation during the quiescent radio pulsar state, failed to detect any signal with a $3\,\sigma$ upper limit of 
$1.4\%$ in the $R$ band \cite{Baglio2016}.

\subsection{1RXS J154439.4--112820}
\label{RXS}

After the recognition of a peculiar variability pattern during the active state of J1023 and J12270, searches for new members of the transitional 
MSPs started, searching for rapid variability in the X--ray light curve of unidentified sources or cataclysmic variables and association with Fermi sources. 
With these characteristics Bogdanov \& Halpern \cite{BogdanovH2015} identified the forth member of the TMSP class in 1RXS J154439.4--112820 
(J154439 in the following). J154439 is the only X--ray source within the $95\%$ error circle of the Fermi source 3FL J1544.6--1125 \cite{Stephen2010}.
J154439 has also been detected during the XMM Slew survey and by Swift/XRT, with flux variations by a factor of $\sim 3$.
Masetti et al. \cite{Masetti2013} provided evidence for the presence of prominent emission lines (H and He) in the $R\sim 18.4$ optical counterpart,
suggesting its identification as a cataclysmic variable.

During an XMM-Newton observation, J154439 showed characteristic rapid variations on a timescale of $\sim 10$ s, passing randomly from $\sim 2$ c s$^{-1}$ 
to $\sim 0.2$ c s$^{-1}$ and back. The overall spectrum is well fit with an absorbed power law model with $\Gamma=1.7$ ($N_H=1.4\times 10^{21}$ cm$^{-2}$,
consistent with the Galactic value). Spectral variations are only marginally evident with indexes of $1.67\pm0.04$ and $1.97\pm0.28$ ($90\%$ confidence 
level) for the high and low states, respectively \cite{BogdanovH2015}. Even if pn data were taken in timing mode, no X--ray periodicity was reported. 
A NuSTAR observation confirmed the spectral parameters over the 0.3--79 keV energy band, leading to a luminosity of $10^{33}$ erg s$^{-1}$ at 
a scale distance of 1 kpc \cite{Bogdanov2016}.
Fast photometric data taken at the MDM Observatory showed fast variability also in the optical counterpart. MDM data and XMM-Newton/OM data 
revealed hints for an orbital modulation at a period of $\sim 5.2-5.4$ hr.

\section{Once upon a time there were a magnetic neutron star interacting with an accretion disc}
\label{sec:5}

The transition from and to the radio pulsar regime observed in the three well studied TMPS J1023, J18245, and J12270 with a good degree
of certainty involves the presence of an accretion disc. In addition to an enhanced emission at all wavelengths, double horned 
emission lines were observed. This testifies that the switch on and off of the radio pulsar mechanism in binary systems can occur, at least,
on timescales of tens of days and it is not a single occurrence in the neutron star lifetime. 

\begin{figure}[h!]
\begin{center}
\includegraphics[scale=0.6]{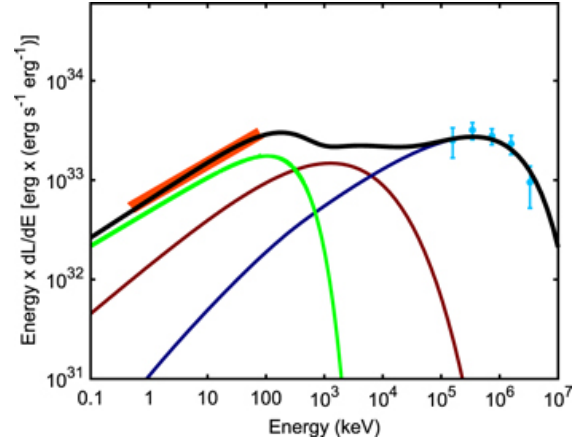}
\end{center}
\caption{
Average spectral energy distribution (SED) observed from PSR J1023+0038 in X--rays (orange strip, from \cite{Tendulkar2014}), and gamma--rays (cyan points, from \cite{Takata2014}), 
evaluated for a distance of 1.37 kpc. The total SED evaluated with Papitto \& Torres \cite{PapittoT2015} modelling is plotted as a black solid line. 
Synchrotron, Syncrotron-Self-Compton, and accretion flow (i.e. the sum of disc and neutron star emission) components are plotted as red, blue and green lines, respectively.}
\label{j1023sed}       
\end{figure}

The inference of the root cause of the sub-luminous disc state is more complicated. 
This sub-luminous state is dominated ($\sim 80\%$ of the time) by the high mode during which X--ray pulsations are observed (e.g. in J1023).
For this high mode Papitto \& Torres (\cite{PapittoT2015}, see also \cite{Bednarek2015}) showed that a scenario based on the propeller can explain 
the observed features at all wavelengths. The spectral energy distribution is characterised a broad emission from X--rays to gamma--rays (see Fig. \ref{j1023sed}). 
Papitto \& Torres \cite{PapittoT2015} interpreted the gamma--ray part as due to the self-synchrotron Compton emission 
that originates at the turbulent boundary between the neutron star magnetosphere propelling the disc inflow at super-Keplerian speed.
The X--ray emission is instead due to the sum of the synchrotron emission that originated from the same magnetospheric region and the 
luminosity emitted by the accretion flow, which however must be inefficient ($\lsim 20\%$ of the conversion of gravitational energy) not to 
exceed the observed one.

Different suggestions have been put forward to explain the transitions among high and low mode (and flaring) and, more importantly, 
why only the three known TMSPs show this puzzling behaviour (and stability).
The flux drops that bring the sources to the low mode are extremely peculiar and are very different from the dips currently observed in 
LMXBs, since there are no associated spectral changes and/or reduction of the soft X--ray flux, excluding absorption from intervening matter.
Apparently, there is also no evidence for an X--ray luminosity dependence on the duration and frequency of flares and low flux mode intervals 
or any correlation between the separation between (and duration of) dips or flares \cite{Bogdanov2015}.

The fate of matter falling onto a magnetic neutron star is determined by the position of three different radii. Two are fixed and depends on the 
neutron star spin period: the corotation radius $r_{\rm cor}=(G\,M_{NS}\,P^2/(4\,\pi^2))^{1/3}$ (where magnetic field lines reach the Keplerian speed) 
and the light cylinder radius $r_{\rm lc}=c\,P/(2\,\pi)$ (with $c$ the light velocity, where the magnetic field lines open, being unable to corotate with 
the neutron star). The truncation of the accretion disc by the magnetosphere occurs at the magnetospheric radius 
$r_{\rm m}=\xi\,(\mu^2/(2\,G\,M\,\mdot^2))^{1/7}$, with $\xi\sim0.5$ \cite{Campana2018} accounting for the disc geometry, 
$\mu=B\,R_{NS}^3$ the dipole magnetic moment,
and $\mdot$ the mass accretion rate at the magnetospheric boundary. If the magnetospheric radius lies within the corotation radius, accretion proceeds
unimpeded, however if $r_{\rm m}>r_{\rm cor}$  the incoming matter experiments a centrifugal force larger than gravity when it gets attached 
to the fields lines at the magnetospheric radius and (ideally) gets propelled out. If the mass accretion rate decreases further, $r_{\rm m}$ expands
further, reaching at some point the light cylinder radius. At this point the magnetic field becomes radiative and matter is expelled further out by 
radiation pressure and relativistic particle wind \cite{Campana1998,Burderi2001}. A radio pulsar can in principle start working again.
The observed luminosity in the sub-luminous state of J1023 and J12270 is relatively low and at the corresponding mass accretion rate (assuming
that all the accreting material arrives at the neutron star surface) implies a magnetospheric radius well outside the corotation radius 
\cite{Archibald2015,Bogdanov2015,Campana2016}, so that the sources should be in the propeller regime.
This poses a problem and suggests intriguing new insights for accretion physics at low luminosities.

The best and well studied example of propeller accretion is probably the accreting white dwarf AE Aquarii. In this system only $\sim 0.3\%$ of the 
incoming material is able to reach the star surface. In AE Aqr most of the soft X--ray luminosity is produced in 
the inflow before ejection or accretion onto the surface \cite{Oruru2012}. Correlation and lag analysis can shed light on the 
emission mechanisms in TMPS. J1023 has been well characterised. Hard (NuSTAR) and soft (XMM-Newton) X--ray emission appear well correlated 
\cite{CotiZelati2018}. X--rays (XMM-Newton) instead do not correlate with UV (Swift/UVOT, \cite{CotiZelati2018}) nor with 
optical ($B$) emission (XMM-Newton/OM, \cite{Bogdanov2015}), unless during the flaring state. This suggests that at least in J1023, as for 
AE Aqr, UV-optical emission comes from the accretion disc and the companion star.

Jaodand et al. \cite{Jaodand2016} were able to measure the overall period evolution of J1023. They found that J1023 is spinning down at a rate that is 
$\sim 30\%$ larger than the pure dipole spin-down rate. This is consistent with the the modelling of Parfrey et al. \cite{Parfrey2017} of AMSP 
magnetospheres
in which the spin-down during propeller has the the same functional form of the pulsar spin down and depends on how the disc inside the light 
cylinder opens some of the closed field lines, leading to an enhancement of the power extracted by the pulsar wind and the spin-down 
torque applied to the pulsar.

One class of scenarios involves a trapped disc.
A propeller may not be able to eject matter so that the inflaming matter stays confined in the innermost part of the flow, trapping the 
magnetospheric radius close to the corotation radius \cite{DAngelo2010,DAngelo2012}.  Pure trapped disc models will not work being 
the oscillation timescale much shorter than what observed \cite{Bogdanov2015}.
The switching between low and high mode might be the result of transitions between a non-accreting pure propeller 
mode and an accreting trapped-disc mode \cite{Archibald2015,Bogdanov2015}.

Alternatively, Campana et al. \cite{Campana2016} (see also \cite{Linaresetal2014}) proposed that high mode is connected to the propeller regime,
whereas the low mode to the  expulsion of the infalling matter by the neutron star pressure, with the neutron star in the radio pulsar state. 
During the high mode in the propeller regime, some matter leaks through the centrifugal barrier and accretes onto the neutron star, as shown 
in magnetohydrodynamical simulations \cite{Romanova2005}, and generates the observed X--ray pulsations.
Detailed spectral modelling is consistent with a radiatively inefficient accretion disc close to the corotation radius during the high mode and 
receding beyond the light cylinder during the low mode. Contemporarily a shock emission sets in. Pulsed emission at a lower amplitude ($\sim 2-3\%$)
and in the soft band only is predicted to occur. This scenario is in agreement with the broadband modelling by \cite{PapittoT2015} and by 
the observed optical polarisation coming from the disc \cite{Baglio2016} and flat radio spectrum \cite{Deller2015}.
The strong anti-correlation among X--ray and radio emission \cite{Bogdanov2017} fits perfectly within this scenario: during the low state matter is expelled
from the system by the relativistic pulsar wind generating in the shock strong radio emission. Parfrey \& Tchekhovskoy \cite{Parfrey2018} showed through 
general-relativistic MHD simulations that this scenario is plausible.

A completely different scenario \cite{Jaodand2016} is motivated by the observation of mode switching in radio pulsars, in which the pulse profile switches 
between two stable profiles. PSR B0943+10 showed that the mode switching in radio is accompanied by simultaneous switches in the X--ray 
band (profile and intensity; \cite{Mereghetti2013,Hermsen2013}). Mode switching has never been observed in any X--ray pulsar so for the 
moment this might remain an intriguing suggestion. 

Optical pulsations in the active X--ray state are puzzling. \cite{Ambrosino2017} showed that cyclotron emission can be ruled out and a rotation-powered 
mechanism might work. A pure (engulfed) radio pulsar however encounters problems in explaining the full phenomenology of J1023 \cite{Campana2016}.
As noted in \cite{Bogdanov2017} J1023 and the other transitional ms pulsars are the only systems for which the neutron star 
rotational energy is comparable to the accretion energy. It might be that the neutron star always looses rotational energy, independently of the fate of 
the accreting matter. Accretion of matter is able to quench pulsed radio emission but pulsed optical emission might be generated. 
In this case we would expect pulsed optical emission to be present independently on the X--ray mode.

\section{Open Questions}
\label{sec:6}

Among all the open questions that we described above, the most puzzling
problem still regards the connection between LMXBs and AMSPs, or MSPs 
in general. It is still a mystery the reason why LMXBs usually do not show 
X--ray pulsations, not even at low mass accretion rates when presumably the 
magnetospheric radius is outside the neutron star surface and inside 
the co-rotation radius, thus allowing matter to be channelled by the
magnetic field. Does it have to do with the magnetic field, that can
be buried inside the neutron star surface by prolonged accretion phases?
Another mystery is why AMSPs, although fast spinning and with a misaligned 
magnetic field do not show pulsed radio or gamma emission (as instead do other 
not accreting MSPs, including transitional MSPs) in quiescence, when accretion 
onto the neutron star decreases by orders of magnitude presumably pushing 
the magnetospheric radius outside the light cylinder radius. Does it have 
to do with the presence of matter around the system that is swept away
by the radiation pressure of the pulsar? Is this enough to justify also
non detections in the gamma-ray band? In other words, it would be important
to understand why a system like J18245 (a.k.a. M28I), swinging 
between the rotation-powered and the accretion-powered phase in such short 
timescales is so rare. Is it because of its relatively large ($\sim 11$ hr) 
orbital period so that the matter outflowing the system is spread in a 
relatively wide orbit?

On the TMSP side, we have a number of open questions that will likely be answered in the next following years.
Why do some sources (J1023 and XSS12270) showed a long (years) accretion-dominated state whereas J18245 showed a proper X--ray outburst?
Why do the accretion-dominated state of these systems so stable (e.g. in J1023)? 
What is the root cause of the mode switching observed during the accretion-dominated state? Are X--ray pulsations detectable 
at a lower level (and/or in a softer energy band) also in the low mode of the active X--ray state?
These are the basic questions that make these systems particularly attractive.

Recently optical pulsations detected in J1023 added a number of additional questions.
What is the mechanism responsible for optical pulsation? Is the accretion-dominated state really powered by accretion?
Can a rotation-powered mechanism remain active (i.e. not inhibited) during an accretion state?

Answering these questions will give important information on the still elusive radio pulsar and accretion emission mechanisms and 
on the evolutive path producing the different types of MSPs known today.

\section{Acknowledgements}

We would like to thank M.C. Baglio, L. Burderi, F. Coti Zelati, P. D'Avanzo, D. de Martino, R. Iaria, A. Papitto, 
N. Rea, A. Riggio, A. Sanna and L. Stella for useful discussions and comments over time.

\input{referenc}

\end{document}

%% file: referenc.tex
%
%
%
%